\documentclass[aos]{imsart}

\RequirePackage{amsthm,amsmath,amsfonts,amssymb}
\RequirePackage[numbers]{natbib}
\RequirePackage[colorlinks,citecolor=blue,urlcolor=blue]{hyperref}
\RequirePackage{graphicx}

\startlocaldefs
\theoremstyle{plain}

\newtheorem{theorem}{Theorem}[section]
\newtheorem{corollary}{Corollary}[section]
\newtheorem{lemma}{Lemma}[section]
\newtheorem{remark}{Remark}[section]
\newtheorem{proposition}{Proposition}[section]

\theoremstyle{remark}

\newtheorem{example}{Example}

\usepackage{hyperref}
\usepackage{color}
\usepackage{amsthm}
\usepackage{amssymb}
\usepackage{amsmath}
\usepackage{epstopdf}
\usepackage{graphicx}
\usepackage{indentfirst}
\usepackage{multirow}
\usepackage{array,arydshln}
\usepackage{subfigure,comment}

\usepackage{stackengine}

\def\OA{{\mbox{\small OA}}}
\def\SOA{{\mbox{\small SOA}}}
\def\CROA{{\mbox{\small CROA}}}

\def\D{{\mbox{\small D}}}

\def\BSOA{{\mbox{\small BSOA}}}
\def\NOA{{\mbox{\small NOA}}}
\def\ROA{{\mbox{\small ROA}}}

\newcommand{\bA}{\textbf{A}}
\newcommand{\bB}{\textbf{B}}
\newcommand{\bD}{\textbf{D}}

\newcommand{\bE}{\textbf{E}}

\newcommand{\ba}{\textbf{a}}
\newcommand{\bb}{\textbf{b}}
\newcommand{\bd}{\textbf{d}}
\newcommand{\bc}{\textbf{c}}

\newcommand{\bff}{\textbf{f}}
\newcommand{\bF}{\textbf{F}}

\newcommand{\bzero}{\textbf{0}}

\newcommand{\bv}{\textbf{v}}
\newcommand{\bu}{\textbf{u}}
\newcommand{\be}{\textbf{e}}

\newcommand\oast{\stackMath\mathbin{\stackinset{c}{0ex}{c}{0ex}{\ast}{\bigcirc}}}

{\begin{list}{}%
         {\setlength{\leftmargin}{#1}}%
         \item[]%
}
{\end{list}}

\endlocaldefs

\begin{document}

\begin{frontmatter}
\title{A New and Flexible Design Construction for Orthogonal Arrays for Modern Applications}
%\title{A sample article title with some additional note\thanksref{t1}}
\runtitle{A New and Flexible Design Construction}
%\thankstext{T1}{A sample additional note to the title.}

\begin{aug}
%%%%%%%%%%%%%%%%%%%%%%%%%%%%%%%%%%%%%%%%%%%%%%%
%% Only one address is permitted per author. %%
%% Only division, organization and e-mail is %%
%% included in the address.                  %%
%% Additional information can be included in %%
%% the Acknowledgments section if necessary. %%
%%%%%%%%%%%%%%%%%%%%%%%%%%%%%%%%%%%%%%%%%%%%%%%
\author[A]{\fnms{Yuanzhen} \snm{He}\ead[label=e1]{heyuanzhen@bnu.edu.cn}},
\author[B]{\fnms{C. Devon} \snm{Lin}\ead[label=e2,mark]{devon.lin@queensu.ca}}
\and
\author[C]{\fnms{Fasheng} \snm{Sun}\ead[label=e3,mark]{sunfs359@nenu.edu.cn}}
%%%%%%%%%%%%%%%%%%%%%%%%%%%%%%%%%%%%%%%%%%%%%%
%% Addresses                                %%
%%%%%%%%%%%%%%%%%%%%%%%%%%%%%%%%%%%%%%%%%%%%%%
\address[A]{School of Statistics,
Beijing Normal University,
\printead{e1}}

\address[B]{Department of Mathematics and Statistics,
Queen's University,
\printead{e2}}

\address[C]{KLAS and School of Mathematics and Statistics,
Northeast Normal University,
\printead{e3}}
\end{aug}

\begin{abstract}
Orthogonal array, a classical and effective tool for collecting data, has been flourished with its applications in modern computer experiments and engineering statistics. Driven by the wide use of computer experiments with both qualitative and quantitative factors, multiple computer experiments, multi-fidelity computer experiments, cross-validation and stochastic optimization,  orthogonal arrays with certain structures have been introduced. Sliced orthogonal arrays and nested orthogonal arrays are examples of such arrays. This article introduces a flexible, fresh construction method which uses smaller arrays and a special structure. The method uncovers the hidden structure of many existing fixed-level orthogonal arrays of given run sizes, possibly with more columns. It also allows fixed-level orthogonal arrays of nearly strength three to be constructed, which are useful as there are not many construction methods for fixed-level orthogonal arrays of strength three, and also helpful for generating Latin hypercube designs  with desirable low-dimensional projections.  Theoretical properties of the proposed method are explored. As  by-products, several theoretical results on orthogonal arrays are obtained.
\end{abstract}



\end{frontmatter}
%%%%%%%%%%%%%%%%%%%%%%%%%%%%%%%%%%%%%%%%%%%%%%
%% Please use \tableofcontents for articles %%
%% with 50 pages and more                   %%
%%%%%%%%%%%%%%%%%%%%%%%%%%%%%%%%%%%%%%%%%%%%%%
%\tableofcontents

\section{Introduction}

With the exponential growth  of computing power, investigators are increasingly
employing computer experiments, which stimulate real-world phenomena or complex systems using mathematical models and solving them using numerical methods such as computational fluid dynamics and finite element analysis, to help understand the respective systems.
The underlying mechanisms of computer experiments are represented and implemented by computer codes \cite{Sacksetal1989, SantnerWilliamsNotz2018}.  They frequently involve  both qualitative and quantitative factors \cite{QianWuWu2008, Huangetal2016,Dengetal2017, Zhangetal2019}. For a given system, investigators are often faced with different computer codes because of the proliferation of mathematical models and numerical methods for their solution. As such, multiple computer experiments for the same system are performed \cite{Yangetal2013}. Computer codes can be also executed at various degrees of accuracy, resulting in multi-fidelity computer experiments \cite{HaalandQian2011, Gohetal2013, WeiXiong2019}.

In contrast to the traditional physical experiments, greater numbers of input variables are involved and larger numbers of runs are employed in computer experiments. However, often only a few of input variables are thought to be of primary importance. To select important input variables, space-filling designs with desirable low-dimensional  projection properties   are commonly used.
Such designs can be generated using orthogonal array based Latin hypercubes  \cite{Owen1992, Tang1993}.
This calls for orthogonal arrays of strength two, three or higher.
To accommodate computer experiments with qualitative and quantitative factors and multiple computer experiments, sliced orthogonal array based Latin hypercubes  were employed \cite{QianWu2009, HeQian2016, HwangHeQian2016, KongAiTsui2018}.   Nested orthogonal array based Latin hypercubes were introduced to choose inputs in multi-fidelity computer experiments \cite{QianTangWu2009}.
The objective of this article is to propose a flexible method that allows orthogonal arrays, sliced orthogonal arrays and nested orthogonal arrays to be constructed.

To balance the run size economy and the higher dimensional projection property, we consider a class of designs, called {\em orthogonal arrays of nearly strength three}, which, for a given run size, enjoy a higher dimensional projection property than orthogonal arrays of strength two and accommodate more factors than those of strength three \cite{Hedayat1986, Hedayat1990}.  This class of designs is in the same spirit as {\em nearly orthogonal arrays} and can be viewed as  their generalization  \cite{WangWu1992, Xu2002, Lekivetzetal2015}. They can be used to generate Latin hypercube designs with desirable low-dimensional projections   which  have been shown beneficial in several problems such as studying design optimality \cite{SunWangXu2019}, and mean and variance estimation which further are useful in numerical integration, uncertainty quantification, and sensitivity analysis \cite{HeQian2014}.

Research on orthogonal arrays has been a central part of design theory for decades and there are extensive and numerous results on their construction, optimality criteria, and analysis. For excellent sources of reference for this topic, see, for example,
\cite{DeyMukerjee1999, HedayatSloaneStufken1999, MukerjeeWu2007, WuHamada2011, XuPhoaWong2009}.  The applications of orthogonal arrays are pervasive in many areas such as  computer experiments, integration, visualization, optimization, and computer science \cite{GopalakrishnanStinson2--6, Owen1992, Hedayat2017}.  The proposed method discovers a new design structure  that  allows orthogonal arrays of strength two, near strength three and exact strength three, resolvable orthogonal arrays, sliced orthogonal arrays and nested orthogonal arrays to be found.  We provide {\em new} orthogonal arrays of strength three, sliced orthogonal arrays and nested orthogonal arrays.  

It has been long since  the concept of resolvable orthogonal arrays was introduced \cite{BoseBush1952}.  There are scarce results on the construction and the use of such arrays. \cite{GuptaNigamDey1982} was the first to discover the use of such arrays in producing orthogonal main-effects plans. Recently, it is shown by
\cite{YangChenLiu2014} and \cite{MukerjeeSunTang2014} that resolvable orthogonal arrays play a key role in constructing sliced Latin hypercubes and nearly orthogonal arrays that are mappable into fully orthogonal arrays, respectively.   A  completely resolvable orthogonal array is used for qualitative factors in marginally coupled designs for  computer experiments with both qualitative and quantitative factors \cite{DengHungLin2015, HeLinSun2017, Heetal2017, HeLinSun2019, YangLinZhouHe2021}.
Notably, \cite{Suen1989} and \cite{HedayatSloaneStufken1999}  are the only work that provided construction for resolvable orthogonal arrays with the former focusing on two-level arrays and the latter on multi-level arrays. The proposed method offers an alternative way to construct multi-level resolvable orthogonal arrays.

The remainder of this article is organized as follows. Section 2 provides necessary definitions, notation and background. Design construction and its use in constructing orthogonal arrays of strength two are given in Section 3. Section 4 presents the applications of the proposed method in constructing a number of classes of orthogonal arrays. Conclusion and discussion are given in Section 5. All the proofs are relegated to the Appendix.

\section{Definitions, Notation and Background}

Consider designs of $n$ runs with $m$ factors of $s$ levels, where $2 \leq s \leq n$.   If for every $n \times t$ submatrix of the design, say $\bD$,  each of
all possible level combinations appears equally often, design $\bD$ is called an {\em orthogonal array} of strength $t$ \cite{HedayatSloaneStufken1999}. We use $\OA(n,m,s,t)$ to denote such a design.
A column is said to be {\em balanced} if each level appears equally often.
A pair of columns is  said to be {\em orthogonal} if they form an orthogonal array of strength two.
Three columns are said to be {\em 3-orthogonal} if they form an orthogonal array of strength three.
In addition,  an $\OA(n,m,s,2)$ is called {\em saturated} if $n = m(s-1)+1$.

Three special classes of orthogonal arrays that are relevant in this article are   resolvable orthogonal arrays, sliced orthogonal arrays and nested orthogonal arrays.
An $\OA(n, m, s, t)$ $\bD$ is called  {\em $\alpha$-resolvable}    if it can be expressed as $\bD=(\bD^T_1, \ldots, \bD^T_{n/(\alpha s)})^T$ such that each of $\bD_1,\ldots,\bD_{n/(\alpha s)}$ is an $\OA(\alpha s, m, s,1)$ \cite{HedayatSloaneStufken1999}, for $t \geq 2$. Such an orthogonal array is denoted by $\ROA(n, m, s, t;\alpha)$. If $\alpha=1$, the corresponding design is known as {\em completely resolvable orthogonal array} and denoted by $\CROA(n,m,s,t)$. For example, the orthogonal array $\bD=\OA(16,3,4,2)$ in Table~\ref{tab:OAs} is a $\CROA(16,3,4,2)$. 

Next, we provide the definitions of both sliced orthogonal arrays and nested orthogonal arrays.
Both definitions involve certain mapping which we call {\em  level-collapsing projection}.  A mapping $\delta(\cdot)$ is called  level-collapsing projection  if the mapping is from a set $S$ with $s$ elements into its own subset and satisfies  (a) $S$ can be divided into $s_0$ parts $S_1, \ldots, S_{s_0}$, with each part having $s/s_0$ elements, and (b) for any two elements $x\in S_i, y\in S_j$, $\delta(x)=\delta(y)$ for $i=j$, and $\delta(x)\neq\delta(y)$ otherwise.  Table~\ref{tab:OAs} provides an example of the level-collapsing projection $\delta(0)=\delta(1) = 0$ and $\delta(2) = \delta(3) = 1$. 
An $\OA(n,m,s,2)$ $\bD$ is called {\em sliced orthogonal array} if its $n$ rows can be partitioned into $v$ subarrays $\bD_1, \bD_2, \ldots, \bD_v$  such that   each $\bD_i$ becomes an $\OA(n_0,m,s_0,2)$ with $n_0 = n/v$ after the $s$ levels in each column of $\bD$ are collapsed to $s_0$ levels according to some level-collapsing projection.
We denote such an array $\SOA(n,m,s,2;v,s_0)$.  Furthermore, if each column in each slice $\bD_i$ of a sliced orthogonal array is balanced, that is, having equal frequency of $s$ levels, it is called {\em balanced sliced orthogonal array} \cite{AiJiangLi2014}. We use $\BSOA(n,m,s,2;v,s_0)$ to denote such an array.
An  $\OA(n,m,s,2)$ $\bD$ is called  {\em nested orthogonal array}  if
 $\bD$ contains a subarray $\bD_0$ which becomes an $\OA(n_0,m,s_0,2)$ after level collapsing each column of $\bD$ according to certain level-collapsing projection \cite{QianAiWu2009}.  We use $\NOA(n,m,s,2;n_0,s_0)$ to denote such an array. As an illustration, Table~\ref{tab:OAs} presents a $\bD=\BSOA(16,3,4,2;4,2)$  which is also an $\NOA(16,3,4,2;4,2)$ with any given $\bD_i$ being the subarray $\bD_0$, for $i=1, 2, 3, 4$.  A sliced orthogonal array is also a nested orthogonal array but the reverse does not hold.

\begin{table}[!ht] \begin{center}
\caption{An orthogonal array $\bD=\OA(16,3,4,2)$ and the level-collapsing projection $\delta$: $\bD$   is also a $\CROA(16,3,4,2)$, a $\BSOA(16,3,4,2;4,2)$ and an $\NOA(16,3,4,2;4,2)$} 
\scalebox{0.9}{
\begin{tabular}{ccccccccc}
\hline
                                 &\multicolumn{4}{c}{$\bD$}  &    &   \multicolumn{3}{c}{$\delta(\bD)$}  \\ \hline
   \multirow{4}{*}{$\bD_1$}      &   $0$     &   $0$      &     $0$   &  &  &  $0$     &   $0$      &     $0$    \\
                                 &   $2$     &   $1$      &    $3$  &  &  &  $1$     &   $0$      &    $1$     \\
                                 &   $1$     &   $3$    &     $2$   &  &  &  $0$     &   $1$      &     $1$    \\
                                 &   $3$   &   $2$      &    $1$    &  &  &  $1$     &   $1$      &    $0$     \\ \hline
   \multirow{4}{*}{$\bD_2$}      &   $2$     &   $2$      &     $2$   &  &  &  $1$     &   $1$      &     $1$    \\
                                 &   $0$     &   $3$    &    $1$    &  &  &  $0$     &   $1$      &    $0$     \\
                                 &   $3$   &   $1$      &     $0$   &  &  &  $1$     &   $0$      &     $0$    \\
                                 &   $1$     &   $0$      &    $3$  &  &  &  $0$     &   $0$      &    $1$    \\ \hline
   \multirow{4}{*}{$\bD_3$}      &   $1$     &   $1$      &     $1$   &  &  &  $0$     &   $0$      &     $0$    \\
                                 &   $3$   &   $0$      &    $2$    &  &  &  $1$     &   $0$      &    $1$     \\
                                 &   $0$     &   $2$      &    $3$  &  &  &  $0$     &   $1$      &    $1$     \\
                                 &   $2$     &   $3$    &    $0$    &  &  &  $1$     &   $1$      &    $0$     \\ \hline
   \multirow{4}{*}{$\bD_4$}      &   $3$   &    $3$   &   $3$   &  &  &  $1$     &   $1$      &   $1$      \\
                                 &   $1$     &   $2$      &    $0$    &  &  &  $0$     &   $1$      &    $0$     \\
                                 &   $2$     &   $0$      &     $1$   &  &  &  $1$     &   $0$      &     $0$    \\
                                 &   $0$     &   $1$      &    $2$    &  &  &  $0$     &   $0$      &    $1$     \\
   \hline
\end{tabular}}\label{tab:OAs}
\end{center}
\end{table}

A relevant concept for orthogonal arrays is difference scheme. An $r\times c$ array with entries from a finite abelian group containing $s$ entries is called a difference scheme if  each vector difference between any two distinct columns of
the array consists of every element from the finite abelian  group
equally often \cite{BoseBush1952}. Such an array is denoted by  $\D(r, c, s)$.

The proposed construction in Section 3 uses two key operators for matrices. Before reviewing the operators, we review the concept of Galois field.  A field is called a Galois field (or finite field) if it  contains a finite number of elements and the operations of the  addition, subtraction, multiplication, and division on its elements satisfy the rules of arithmetic  (\cite{HedayatSloaneStufken1999}). The order of a Galois field is defined to be the number of elements, and must be a prime power.  
Let $GF(s)$ denote the Galois field of order $s$, and throughout we let $GF(s) =\{\alpha_0, \alpha_1, \ldots, \alpha_{s-1}\}$ with $\alpha_0=0$ and $\alpha_1=1$. 
If $s=p$ is a prime,  $GF(p)=\{0, 1, \ldots, p-1\}$. If $s=p^u$ for $u \ge 2$ and a prime $p$, $GF(p^u)=\{ a_0+a_1x+\cdots + a_{u-1} x^{u-1} : a_0, \ldots, a_{u-1} \in GF(p) \}$. That is, the elements of $GF(p^{u})$ are the polynomials with the degree less than $u$ and the coefficients from $GF(p)$. 
The addition of $GF(p^{u})$ is the ordinary polynomial addition with the coefficients modulo $p$,  and the multiplication is the ordinary polynomial multiplication and then modulo a given irreducible polynomial  of degree $u$, where an irreducible polynomial   is a polynomial that cannot be factored into the product of two non-constant polynomials. For examples of the addition and multiplication of a Galois field, readers are referred to \cite{HedayatSloaneStufken1999}.

Let $\bA=(a_{ij})$ be an $n_1 \times m_1$ matrix and $\bB$ be an $n_2 \times m_2$ matrix, where both matrices have entries from the Galois field $GF(s)$. Throughout, we let $GF(s) = \{ \alpha_0, \alpha_1, \ldots, \alpha_{s-1}\}$ with $\alpha_0 = 0$ and $\alpha_1 = 1$. The {\em Kronecker sum} of $\bA$ and $\bB$ is an $(n_1n_2) \times (m_1m_2)$ matrix given by
\begin{equation}\label{eq:ks}
\bA \oplus \bB= [{\bB}^{a_{ij}}]_{1\leq i\leq n_1, 1\leq j\leq m_1},
\end{equation}
\noindent where ${\bB}^{a_{ij}}=(\bB+ a_{ij})$ is an $n_2 \times m_2$  matrix with $+$ representing the addition in a field.  If we partition   the rows of $\bB$ into $n_1$ matrices,  $\bB_{i}$'s, and let $\ba_i$ be the $i$th row of $\bA$,   the {\em generalized Kronecker sum} of $\bA$ and $\bB$ is defined as 
\begin{equation}\label{eq:gks}
\bA \oast \bB = [\ba_{i} \oplus \bB_{i}]_{1 \leq i \leq n_1}=
  \left(
  \begin{array}{c}
    \ba_{1} \oplus \bB_{1} \\
    \vdots \\
    \ba_{n_1} \oplus \bB_{n_1}
  \end{array}
\right),
\end{equation}
where  the operator $\oplus$ is given in (\ref{eq:ks}).  Note that the number of runs in $\bA \oast \bB$ is the same as that of $ \bB$ and the number of factors in $\bA \oast \bB$ is $m_1m_2$, the product of the number of factors in $\bA$ and the number of factors in $\bB$. For example, let $\bA = (0,1,2)^T$ and 
\begin{center}
$\bB = \left( 
\begin{array}{r}
\bB_1\\
\bB_2\\
\bB_3
\end{array}
\right)  =\left( \begin{array}{rrrr}
0 & 0 & 0 & 0 \\
0 & 1 & 1 & 2\\
0 & 2 & 2 & 1\\
\hdashline 
1 & 0 & 1 & 1\\
1 & 1 & 2 & 0 \\
1 & 2 & 0 & 2 \\
\hdashline 
2 & 0 & 2 & 2 \\
2 & 1 & 0 & 1\\
2 & 2 & 1 & 0
\end{array} 
\right)$, then we have 
$\bA \oast \bB =  \left( \begin{array}{rrrr}
0 & 0 & 0 & 0 \\
0 & 1 & 1 & 2\\
0 & 2 & 2 & 1\\
2 & 1 & 2 & 2\\
2 & 2 & 0 & 1 \\
2 & 0 & 1 & 0 \\
1 & 2 & 1 & 1 \\
1 & 0 & 2 & 0\\
1 & 1 & 0 & 2
\end{array} 
\right)$.
\end{center}

\section{Design Construction and Preliminary Results}

This section introduces a new construction and presents some preliminary results for the later development. The proposed construction uses  an $n_1 \times m_1$ matrix $\bA$  and $n_1$ matrices $\bB_1, \ldots, \bB_{n_1}$ each of size $n_2 \times m_2$, where the entries of $\bA$ and $\bB_{i}$'s are from the Galois field $GF(s)
=\{ \alpha_0, \alpha_1, \ldots, \alpha_{s-1}\}$. Assume $s$ is a prime power throughout.
Denote the rows of $\bA$ by $\ba_1,\ldots,\ba_{n_1}$, and stack  $\bB_1, \ldots, \bB_{n_1}$ row by row and denote the resulting matrix by $\bB =  (\bB_{1}^T, \ldots, \bB_{n_1}^T )^T$. We define $s+1$ arrays $\bD_1, \bD_2, \ldots, \bD_{s+1}$ as follows. For $\alpha_g \in GF(s)$  and $g=1,\ldots,s-1$, define
\begin{equation}\label{eq:construction2}
\bD_{g}  =  \bA \oast ( \alpha_g * \bB) = \left[ \begin{array}{c}
                                                \ba_1 \oplus  ( \alpha_g * \bB_1)\\
                                                \vdots \\
                                                \ba_{n_1} \oplus  ( \alpha_g * \bB_{n_1})
                                              \end{array}\right ],
\end{equation}
and  
\begin{equation}\label{eq:construction2s}
\bD_{s}  =  \bzero_{n_1} \oast   \bB = \left[ \begin{array}{c}
                                                0 \oplus     \bB_1\\
                                                \vdots \\
                                                0 \oplus   \bB_{n_1}
\end{array}\right ],
\end{equation}
\noindent as well as
\begin{equation}\label{eq:construction2splus1}
\bD_{s+1} = \bA \oplus \bzero_{n_2},
\end{equation}
\noindent where  $*$ in (\ref{eq:construction2}) represents the multiplication in a field and $\bzero_{n_2}$ denotes a column vector with $n_2$ zeros.    Now obtain an array
\begin{equation}\label{eq:E}
\bE=
    \left [\bD_{1},\bD_2, \ldots,  \bD_{s+1} \right ],
\end{equation}
\noindent where each $\bD_g$ is an $(n_1n_2) \times (m_1m_2) $ array by the definition of the generalized Kronecker sum, $g = 1, 2, \ldots, s-1$, $\bD_{s}$  is an $(n_1n_2) \times m_2$ array and $\bD_{s+1}$ is an  $(n_1n_2) \times m_1$ array, and thus $\bE$ is an $(n_1n_2) \times [(s-1)m_1m_2+m_1+m_2]$ array.  

To the best of our knowledge, the construction of $\bE$ in (\ref{eq:E}) is a new way of constructing designs.
The construction is flexible as it can be used to construct different types of orthogonal arrays with different  choices of $\bA$ and $\bB_i$'s.
It produces a rich, vast class of designs with non-isomorphic $\bA$'s, non-isomorphic $\bB_i$'s, as well as isomorphic $\bB_{i}$'s up to column permutations and/or level relabeling of one or more columns (For example, \cite{XuChengWu2004} demonstrates that the level relabeling of columns improves the design properties). In Section 4.1, we illustrate this strategy to provide new orthogonal arrays of strength three. 

To prepare for the detailed and quantitative illustration of the use of the proposed construction, we provide some preliminary results summarized in Lemmas~\ref{lem:2} and \ref{lem:3}.
Some notations are in order.
Let $\ba$, $\bc$ and $\be$ be a column vector of length $n_1$, respectively, whose entries are from $GF(s)$. For $i=1,\ldots,n_1$, let $\bb_{i}$, $\bd_{i}$  and $\bff_{i}$ be a column vector of length $n_2$.
The entries of $\bb_{i}$'s, $\bd_{i}$'s and $\bff_{i}$'s are also from $GF(s)$.
Stacking all $\bb_i$'s row by row, we obtain $\bb = (\bb_{1}^T, \ldots, \bb_{n_1}^T)^T$. Similarly,  obtain $\bd = (\bd_{1}^T, \ldots, \bd_{n_1}^T)^T$ and $\bff = (\bff_{1}^T, \ldots, \bff_{n_1}^T)^T$.
Lemma~\ref{lem:2} provides the sufficient conditions for two columns to be orthogonal  when one column is obtained by the generalized Kronecker sum of $\ba$ and $\bb$ and the other column is obtained by the generalized Kronecker sum of $\bc$ and $\bd$.

\begin{lemma}\label{lem:2} Given $\ba, \bb, \bc, \bd$ defined above, the array $({\ba\oast\bb, \bc\oast\bd})$ is an $OA(n_1n_2, 2, s, 2)$ if one of the following conditions is satisfied, for $i=1,\ldots, n_1$ and $\alpha\in GF(s)$,%\\[-0.33in]
\begin{itemize}
  \item[(i)]   $(\bb_{i}, \bd_{i})$ is an $\OA(n_2, 2, s, 2)$;%\\[-0.33in]
  \item[(ii)] $\bb_i$ is balanced, $\bd = \alpha\bb$, and   one of the following four conditions is satisfied:
                 $(1)$ $\alpha=1$ and $(\ba, \bc)$ is a $\D(n_1, 2, s)$;
                 $(2)$ $(\ba,\bc)$ is an $\OA(n_1,2, s, 2)$; $(3)$ $\bc=\bzero$, $\alpha\neq 0$, and $\ba$ is balanced;
                 and $(4)$ $\ba=\bc$, $\alpha\neq 1$, and $\ba$ is balanced; and%\\[-0.33in]
  \item[(iii)] $\bd = \bzero$, and one of the following two conditions is satisfied: $(1)$ $\bb_i$ and $\bc$ are
                 balanced; and $(2)$ $\bb=\bzero$ and $(\ba, \bc)$ is an $\OA(n_1, 2, s, 2)$.
\end{itemize}
\end{lemma}
 
Lemma~\ref{lem:3} below provides  the sufficient conditions for three columns each obtained by the generalized Kronecker sum to be 3-orthogonal.

\begin{lemma}\label{lem:3}
Given $\ba, \bb, \bc, \bd, \be, \bff$ defined above, the array ($\ba\oast\bb$, $\bc\oast\bd$, $\be\oast\bff$) is an $\OA(n_1n_2, 3, s, 3)$ if one of the following conditions is satisfied, for $i=1,\ldots, n_1$ and $\alpha, \beta\in GF(s)$,%\\[-0.33in]
\begin{itemize}
  \item[(i)]  $(\bb_i, \bd_i,\bff_i)$ is an $\OA(n_2, 3, s, 3)$;%\\[-0.33in]
  \item[(ii)]  $(\bb_i,\bd_i)=\OA(n_2, 2, s, 2)$, $\bff= \alpha\bb$, 
                 and $(\be, \alpha\ba)=\D(n_1, 2, s)$; %\\[-0.33in]
  \item[(iii)]  $ \bd=  \beta\bb$, $\bff= \alpha\bb$, $\bb_i$ is balanced, $(\ba, \bc)=\OA(n_1, 2, s, 2)$
                 and one of the following three conditions is satisfied:
                 $(1)$ $\be=\ba$ and $\alpha\neq 1$; or $(2)$ $\be=\bzero$ and $\alpha\neq 0$; and (3) $(\ba,\bc,\be) = \OA(n_1,3,s,3)$; and%\\[-0.33in]
  \item[(iv)] $\bb = \bd = \bzero$, $\bff_i$ is balanced, and $(\ba, \bc) = \OA(n_1,2,s,2)$.
\end{itemize}
\end{lemma}

We now present one key result of  the proposed construction (\ref{eq:E}). That is, for a prime power  $s$, if $\bA$ and $\bB_{i}$'s in (\ref{eq:E}) are $s$-level orthogonal arrays of strength two,
so is the resulting design $\bE$.

\begin{theorem}\label{thm:e}
If $\bA =   \OA(n_1,m_1,s,1)$ for $m_1=1$ or $\OA(n_1,m_1,s,2)$ for $m_1>1$, and $\bB_{i} = \OA(n_2,m_2,s,2)$, for $i=1,\ldots,n_1$ and a prime power $s$, then design $\bE$ in  (\ref{eq:E}) is an $\OA(n_1n_2,(s-1)m_1m_2+m_1+m_2,s,2)$.
\end{theorem}

Theorem~\ref{thm:e} indicates that an $s$-level orthogonal array of strength two can be constructed
using only $s$-level orthogonal arrays of strength two with smaller sizes. To the best of our knowledge, such a construction is fundamentally different from the existing methods such as direct constructions or those using difference schemes and smaller orthogonal arrays \cite{HedayatSloaneStufken1999}.  To help understand Theorem~\ref{thm:e},  one may think the columns of $\bE$ are derived in the following way. Arrays $\bA$ and $\bB$ accommodate $m_1$ and $m_2$ factors, respectively.  The columns of $\bD_{s+1}$ in (\ref{eq:construction2splus1}) and those of $\bD_{s}$ in (\ref{eq:construction2s})  can be thought of being derived from the  main effects of factors in $\bA$ and  the main effects of factors in $\bB$, respectively. The columns of $\bD_1,\ldots, \bD_{s-1}$ can be viewed as representing the interaction effects of the factors in $\bA$ and those in $\bB$.   For example, for $s=2$, $\bD_1$  constitutes the two-factor interactions between the factors in $\bA$ and those in $\bB$; for $s=3$, $\bD_1 = \bA \oast \bB$ and $\bD_2 = \bA \oast (2*\bB)$ amount to the linear-by-linear interactions and the
linear-by-quadratic interactions, respectively, which together constitute the interaction effects between $\bA$ and $\bB$.

Example \ref{ex:const1} is an illustration of the use of Theorem~\ref{thm:e}.

\begin{example} \label{ex:const1}
Consider constructing three-level orthogonal arrays of 81 runs and strength 2. It is well-known that an $\OA(81,40, 3,2)$ can be constructed via the Rao-Hamming construction  \cite{HedayatSloaneStufken1999}.  The proposed construction  (\ref{eq:E})  can also provide $\OA(81,40, 3,2)$'s using two different ways: (1) using $\bA= \OA(3,1,3,1)$ and $\bB_i = \OA(27,13,3,2)$ for $i=1, 2, 3$, where $\bB_i$ can be any design that is isomorphic to any of the 68 non-isomorphic $\OA(27,13,3,2)$'s given on the website  \cite{EendebakSchoen2017}; and (2) using $\bA = \OA(9,4,3,2)$ and $\bB_i = \OA(9,4,3,2)$ for $i=1,\ldots,9$, where $\bA$ and $\bB_i$'s can be any design that is isomorphic to the unique $\OA(9,4,3,2)$ given on the website \cite{Sloane2004}.
\end{example}

Next,  we apply Theorem~\ref{thm:e} to obtain some specific classes of orthogonal arrays.   Corollaries~\ref{col:e2level_1} - \ref{col:e2} summarize that the proposed construction  can provide two-level {\em saturated} orthogonal arrays whose run sizes are multiples of 8, $s$-level {\em saturated} orthogonal arrays of  $s^u$ runs, and $s$-level orthogonal arrays of $2s^u$ runs, where $u\geq 2$ is an integer and $s$ is  a prime power.  It shall be noted that these three corollaries also hold for a trivial case that $\bA$ is a column vector $(\alpha_0,\ldots,\alpha_{s-1})^T$  for obtaining $s$-level orthogonal arrays.

\begin{corollary}\label{col:e2level_1}
If $\bA = \OA(n_1,n_1-1,2,2)$ and $\bB_{i} = \OA(n_2,n_2-1,2,2)$, for $i=1,\ldots,n_1$, then design $\bE$ in  (\ref{eq:E})  is an $\OA(n_1n_2,n_1n_2-1,2,2)$.
\end{corollary}

\begin{corollary}\label{col:e1}
If $\bA = \OA(s^{k_1},(s^{k_1}-1)/(s-1),s,2)$ and $\bB_{i} = \OA(s^{k_2},(s^{k_2}-1)/(s-1),s,2)$, for $i=1,\ldots,s^{k_1}$ and a prime power $s$, then design $\bE$ in  (\ref{eq:E})  is an $\OA(s^{k_1+k_2},(s^{k_1+k_2}-1)/(s-1), s,2)$.
\end{corollary}

\begin{corollary}\label{col:e2}
If $\bA = \OA(s^{k_1},(s^{k_1}-1)/(s-1),s,2)$ and $\bB_{i} = \OA(2s^{k_2},2(s^{k_2}-1)/(s-1)-1,s,2)$, for $i=1,\ldots,s^{k_1}$ and a prime power $s>2$, then design $\bE$ in  (\ref{eq:E})  is an $\OA(2s^{k_1+k_2},2(s^{k_1+k_2}-1)/(s-1)-1-s(s^{k_1}-1)/(s-1), s,2)$.
\end{corollary}

\begin{remark}
The family of $\OA(2s^{k},2(s^{k}-1)/(s-1)-1,s,2)$ can be obtained via the use of difference schemes for a prime power $s$ and $k \geq 2$ (Theorem 6.40 of \cite{HedayatSloaneStufken1999}). Corollary~\ref{col:e2} reveals that the construction  (\ref{eq:E})  provides $s(s^{k_1}-1)/(s-1)$ fewer columns than the difference schemes method. For a fixed $k=k_1+k_2$, to maximize the number of columns in $\bE$, $k_1 = 1$ should be chosen.
\end{remark}

Table~\ref{table:oa} provides  examples of orthogonal arrays of strength two constructed via (\ref{eq:E}). The list is not meant to be exhaustive.  Table~\ref{table:oa} also lists the maximal number $m^*$ of columns in an orthogonal array that is obtained by the approaches in the literature. It reveals that in all cases except run sizes being $2s^k$ with a prime power $s$,  the proposed construction  (\ref{eq:E})  provides orthogonal arrays with the same $m^*$. We provide some results on  the maximal number of columns in $\bE$ in  (\ref{eq:E})  for given values of $n$, $s$ and $t$ in the Supplementary Material (\cite{HeLinSun2021}).

\begin{table}[!th]
\caption{Examples of orthogonal arrays of strength two obtained by  (\ref{eq:E}) }
\begin{center}
\scalebox{0.9}{
\begin{tabular}{rrrrrrrrl}
\hline
$s$ & $n$ & $m$ & $n_1$ & $m_1$ & $n_2$ & $m_2$ &$m^*$ & Method\\
\hline
2 & 8 & 7 & 2 & 1 & 4 & 3 & 7 & Corollaries~\ref{col:e2level_1} and~\ref{col:e1}\\
2 & 16 & 15 & 2 & 1 & 8 & 7 & 15 & Corollaries~\ref{col:e2level_1} and~\ref{col:e1}\\
2 & 16 & 15 & 4 & 3 & 4 & 3 & 15 & Corollaries~\ref{col:e2level_1} and~\ref{col:e1}\\
2 & 24 & 23 & 2 & 1 & 12 & 11 & 23 & Corollary~\ref{col:e2level_1}\\
2 & 32 & 31 & 2 & 1 & 16 & 15 & 31 & Corollaries~\ref{col:e2level_1} and~\ref{col:e1}\\
2 & 32 & 31 & 4 & 3 & 8 & 7 & 31 & Corollaries~\ref{col:e2level_1} and~\ref{col:e1}\\
2 & 40 & 39 & 2 & 1 & 20 & 19 & 39 & Corollary~\ref{col:e2level_1}\\
2 & 48 & 47 & 2 & 1 & 24 & 23 & 47 & Corollary~\ref{col:e2level_1}\\
2 & 48 & 47 & 4 & 3 & 12 & 11 & 47 & Corollary~\ref{col:e2level_1}\\
2 & 56 & 55 & 2 & 1 & 28 & 27 & 55 & Corollary~\ref{col:e2level_1}\\
2 & 64 & 63 & 2 & 1 & 32 & 31 & 63 & Corollaries~\ref{col:e2level_1} and~\ref{col:e1}\\
2 & 64 & 63 & 4 & 3 & 16 & 15 & 63 & Corollaries~\ref{col:e2level_1} and~\ref{col:e1}\\
2 & 64 & 63 & 8 & 7 & 8 & 7 & 63 & Corollary~\ref{col:e2level_1}\\
2 & 72 & 71 & 2 & 1 & 36 & 35 & 71 & Corollary~\ref{col:e2level_1}\\
2 & 80 & 79 & 2 & 1 & 40 & 39 & 79 & Corollary~\ref{col:e2level_1}\\
2 & 80 & 79 & 4 & 3 & 20 & 19 & 79 & Corollary~\ref{col:e2level_1}\\
2 & 88 & 87 & 2 & 1 & 44 & 43 & 87 & Corollary~\ref{col:e2level_1}\\
2 & 96 & 95 & 2 & 1 & 48 & 47 & 95 & Corollary~\ref{col:e2level_1}\\
2 & 96 & 95 & 4 & 3 & 24 & 23 & 95 & Corollary~\ref{col:e2level_1}\\
3 & 27 & 13 & 3 & 1 & 9 & 4 & 13 & Corollary~\ref{col:e1}\\
3 & 54 & 22 & 3 & 1 & 18 & 7 & 25 & Corollary~\ref{col:e2}\\
3 & 81 & 40 & 3 & 1 & 27 & 13 & 40 & Corollary~\ref{col:e1}\\
3 & 81 & 40 & 9 & 4 & 9 & 4 & 40 & Corollary~\ref{col:e1}\\
3 & 162 & 76 & 3 & 1 & 54 & 25 & 79 & Corollary~\ref{col:e2}\\
4 & 64 &  21 & 4 & 1 & 16 & 5 & 21 & Corollary~\ref{col:e1}\\
4 & 128 &  37 & 4 & 1 & 32 & 9 & 41 & Corollary~\ref{col:e2}\\
4 & 256 &  85 & 4 & 1 & 64 & 21 & 85 & Corollary~\ref{col:e1}\\
4 & 256 &  85 & 16 & 5 & 16 & 5 & 85 & Corollary~\ref{col:e1}\\
5 & 125 & 31 & 5 & 1 & 25 & 6 & 31 & Corollary~\ref{col:e1}\\
5 & 250 & 56 & 5 & 1 & 50 & 11 & 61 & Corollary~\ref{col:e2}\\
5 & 625 & 156 & 5 & 1 & 125 & 31 & 156 & Corollary~\ref{col:e1}\\
5 & 625 & 156 & 25 & 6 & 25 & 6 & 156 & Corollary~\ref{col:e1}\\
7 & 343 & 57 & 7 & 1 & 49 & 8 & 57 & Corollary~\ref{col:e1}\\
7 & 686 & 106 & 7 & 1 & 98 & 15 &113 & Corollary~\ref{col:e2}\\
7 &2401 & 400 & 7 & 1 & 343 & 57 & 400 & Corollary~\ref{col:e1}\\
7 &2401 & 400 &49 & 8 & 49 & 8 & 400 & Corollary~\ref{col:e1}\\
8 & 512 & 73 & 8 & 1 & 64 & 9 & 73 & Corollary~\ref{col:e1}\\
8 & 1024 & 137 & 8 & 1 & 128 & 17 & 145 & Corollary~\ref{col:e2}\\
8 & 4096 & 585 & 8 & 1 & 512  & 73 & 585 & Corollary~\ref{col:e1}\\
8 & 4096 & 585 & 64 & 9 & 64 & 9 & 585 & Corollary~\ref{col:e1}\\
9 & 729 & 91 & 9 & 1 & 81 & 10 & 91 & Corollary~\ref{col:e1}\\
9 & 1458 & 172 & 9 & 1 & 162 & 19 & 181 & Corollary~\ref{col:e2}\\
9 & 6561 & 820 & 9 & 1 & 729 & 91 & 820 & Corollary~\ref{col:e1}\\
9 & 6561 & 820 & 81 & 10 & 81 & 10 & 820 & Corollary~\ref{col:e1}\\
\hline
\end{tabular}}\label{table:oa}
\end{center}
\end{table}

\section{Applications}

The versatility of the proposed construction is demonstrated in this section by showing that it can be used to construct many  orthogonal arrays including those of strength three, resolvable orthogonal arrays, balanced sliced orthogonal arrays, and nested orthogonal arrays.  New orthogonal arrays of strength three, balanced sliced orthogonal arrays and nested orthogonal arrays are obtained. 
The basic idea is to couple certain types of orthogonal arrays with orthogonal arrays of strength two to generate orthogonal arrays of the same types but with larger run sizes and higher dimensionality.  Another key result in this section is, by removing some columns from orthogonal arrays of strength two in the previous section, we get designs that have a very large proportion of  3-orthogonal columns while the numbers of their columns are much larger than the corresponding orthogonal arrays of strength three.  

\subsection{Construction of orthogonal arrays of strength three}
 
In (\ref{eq:construction2})  and (\ref{eq:construction2s}), we let $\bA = \OA(s,1,s,1)$ or $\bA = \OA(s^2,2,s,2)$ and $\bB_i$'s be orthogonal arrays of strength three. Propositions~\ref{prop:ne} and \ref{prop:ne3} summarize the detailed results on the use of the proposed construction for orthogonal arrays of strength three.
For a given run size and number of column in an orthogonal array of strength three, there are two different ways of construction via the proposed method. For example, an $\OA(81,8,3,3)$ can be obtained using either $\bA = (0,1,2)^T, \bB_{i} = \OA(27,4,3,3)$ for $i=1,2,3$ in Proposition~\ref{prop:ne}
or $\bA = \OA(9,2,3,2), \bB_{i} = \OA(9,2,3,2)$ for $i=1,\ldots,9$ in Proposition~\ref{prop:ne3}.

\begin{proposition}\label{prop:ne}
If  for $i=1,\ldots,s$ and a prime power $s$, $\bA$ is an $\OA(s,1,s,1)$ and $\bB_{i}$ is an $OA(n_2,m_2,s,2)$ for $m_2 =2$ or an $\OA(n_2,m_2,s,3)$ for $m_2 \geq 3$,  then for $1 \leq g \neq g' \leq s$, $(\bD_{g},\bD_{g'})$  is an $\OA(sn_2,2m_2,s,3)$, where $\bD_{g}$ is defined 
in (\ref{eq:construction2}) for $g=1,\ldots,s-1$, and  $\bD_{s}$ is defined in (\ref{eq:construction2s}).
\end{proposition}

\begin{proposition}\label{prop:ne3}
If for $i=1,\ldots,s^2$ and a prime power $s$, $\bA$ is an $\OA(s^2,2,s,2)$, and $\bB_{i}$ is an $OA(n_2,m_2,s,2)$ for $m_2 =2$ or an $OA(n_2,m_2,s,3)$ for $m_2 \geq 3$, then \begin{itemize}
\item[(i)] for $1 \leq g \neq g' \leq s-1$, $(\bD_{g},\bD_{g'})$   is an $\OA(s^2n_2,4m_2,s,3)$, and
\item[(ii)] for $1 \leq g \leq s-1$, $(\bD_{g},\bD_{s})$  is an $\OA(s^2n_2,3m_2,s,3)$,
\end{itemize}
\noindent where  $\bD_{g}$  is defined in (\ref{eq:construction2}) for $g=1,\ldots,s-1$, and  $\bD_{s}$ is defined in (\ref{eq:construction2s}).
\end{proposition}

Proposition~\ref{prop:ne} follows by parts (i) and (ii) of Lemma \ref{lem:3}. Proposition  \ref{prop:ne3} requires conditions (i), (ii), (iii.1), and (iii.2) in Lemma \ref{lem:3}.
Both propositions  are powerful results in that by repeated applications of these results, many infinite series of orthogonal arrays of strength three can be obtained.  For example, by Proposition~\ref{prop:ne},   using an $\OA(81,10,3,3)$ for $\bB_{i}$ yields $\OA(243,20,3,3)$'s which in turn can be used to obtain  a pool of $\OA(729,40,3,3)$'s and so on. 
The general result is given in Theorem~\ref{thm:oa3}. In addition, we derive some new results on  lower bounds of the maximal number of columns in certain orthogonal arrays of strength three in the Supplementary Material (\cite{HeLinSun2021}).

\begin{theorem}\label{thm:oa3}
If  there exists an $\OA(n_2,m_2,s,3)$ for a prime power $s$,  then $\OA(n_2s^k,2^km_2,s,3)$'s for $s \geq 2$ can be constructed.
\end{theorem}

Examples of the application of Theorem~\ref{thm:oa3} are obtaining orthogonal arrays of strength three including an $\OA(2^k,2^{k-1},2,3)$, an $\OA(3\cdot 2^k,3\cdot 2^{k-1},2,3)$, an $\OA(2 \cdot 3^k,5\cdot 2^{k-3},3,3)$, an $\OA(3^{k+1},10\cdot 2^{k-3},3,3)$,
 and an $\OA(4^{k+1},17\cdot 2^{k-3},4,3)$, for $k\geq 3$, with   small orthogonal arrays such as an $\OA(8,4,2,3)$,  an $\OA(24,12,2,3)$,  an $\OA(54,5,3,3)$, an $\OA(81,10,3,3)$,
 and an $\OA(256,17,4,3)$.   Next,  we give a detailed example of using the proposed construction to provide new orthogonal arrays of strength three   in Example~\ref{ex:oa3_243}.

\begin{example}\label{ex:oa3_243}
Consider constructing $\OA(243, 20, 3, 3)$'s for which the only available one is provided in \cite{Sloane2004}.  We show how to use Proposition~\ref{prop:ne} to construct new $\OA(243, 20, 3, 3)$'s using an $\OA(81, 10, 3, 3)$ which is unique and available at 
\cite{EendebakSchoen2017}.  Let $\bB_0$ be the $\OA(81, 10, 3, 3)$. 
In Proposition~\ref{prop:ne}, let $\bA = (0,1,2)^T$,  $\bB_1, \bB_2, \bB_3$ be obtained from $\bB_0$ by column permutations, and the resulting design $\bD = (\bD_1, \bD_2)$. By using different column permutations for $\bB_1, \bB_2, \bB_3$, we obtain many different $\bD$'s some of which are isomorphic while some are non-isomorphic to each other and to the  $\OA(243, 20, 3, 3)$ in \cite{Sloane2004}. To save the space, we only list a few ways of obtaining non-isomorphic $\OA(243, 20, 3, 3)$'s in Table~\ref{tab:oa3_243} with the elements being the column permutation of $\bB_0$ for the respective $\bB_1, \bB_2$, and  $\bB_3$. 

\begin{table}[h]
\begin{center}
\caption{Examples of column permutations of $\bB_0$  to obtain $\bB_1, \bB_2, \bB_3$ for non-isomorphic $\OA(243, 20, 3, 3)$'s}
\begin{tabular}{    l l   l l l  }
\hline
\multicolumn{2}{c}{ Design 1}  & $\quad $ & \multicolumn{2}{c}{ Design 2}  \\
\cline{1-2}  \cline{4-5} 
$\bB_1$ &   2    10     4     5     3     8     7     1     6     9    &  & $\bB_1$ &     8     9     2    10     3     7     4     1     5     6\\
  $\bB_2$ &     5     2     1     7     6     8     9    10     3     4 & &  $\bB_2$ &     7     4     8     2     6    10     9     3     1     5\\
  $\bB_3$ &   5     4    10     1     8     6     9     3     2     7 & & $\bB_3$ &     6     1     9     4    10     3     8     7     2     5\\
  &&&& \\ 
\cline{1-2}  \cline{4-5} 
  \multicolumn{2}{c}{ Design 3}  & $\quad $ & \multicolumn{2}{c}{ Design 4}  \\
  \cline{1-2}  \cline{4-5} 
$\bB_1$ &  9     2     1     6    10     8     3     7     4     5   &  & $\bB_1$ &    3     7     1     9     8     4    10     5     6     2\\
  $\bB_2$ &  1     5     7     3     4     6     8     9    10     2 & &  $\bB_2$ &     6    10     5     3     4     1     9     8     7     2 \\
  $\bB_3$ &   9     2     5     3    10     1     7     6     4     8 & & $\bB_3$ &   6     8     9     7    10     3     5     4     2     1\\
\hline  
\end{tabular}\label{tab:oa3_243} 
\end{center}
\end{table} 
\end{example}

\subsection{Construction of orthogonal arrays of near strength three}

We introduce orthogonal arrays of near strength three. These designs are not of exact strength three, but of strength two. They have  the feature that a large proportion
of all possible three-column combinations is 3-orthogonal (see the definition of 3-orthogonality in Section 2).   By slightly relaxing the exact strength-three orthogonality, these designs are shown to accommodate significantly more columns than orthogonal arrays of strength three with the same run sizes. To measure how close these designs are to orthogonal arrays of strength three in terms of 3-orthogonality, we define the following quantity for an orthogonal array $\bD$ with $m$ columns,  \begin{equation}\label{eq:p}
 p(\bD) = \frac{r}{{m \choose 3}},
\end{equation}
\noindent where $r$ is the number of three distinct columns that are 3-orthogonal. Note that $0 \leq p \leq 1$. An orthogonal array of strength three has $p=1$. A larger value of $p$ is preferred.

\begin{proposition}\label{prop:ne2}
If $\bA = \OA(s,1,s,1)$, $\bB_{i} = \OA(n_2,m_2,s,3)$, for $i=1,\ldots,s$ and a prime power $s$, and $\bF$ is obtained by excluding the last column of $\bE$ in (\ref{eq:E}), then $\bF$ is an $\OA(n_2s, sm_2, s,2)$ and
\begin{equation}\label{eq:pf1}
p(\bF) =  1- \frac{m_2{s \choose 3}}{{sm_2 \choose 3}} = 1 - \frac{(s-1)(s-2)}{(sm_2-1)(sm_2-2)}.
\end{equation}
\end{proposition}

Proposition \ref{prop:ne2} can be easily verified by noting that $\bF = ( \bD_{1},\ldots,\bD_{s})$ with $\bD_{g}$ in (\ref{eq:construction2}) for $g =1,\ldots,s-1$ and  $\bD_{s}$ as in (\ref{eq:construction2s}),  and that any three columns that are not 3-orthogonal  satisfy that each of them is from one of $\bD_{i},\bD_{j},\bD_{k}$ with distinct values of $i,j,k$ and obtained by  the generalized Kronecker sum of the same column of $\bB$. The last column of $\bE$ is removed to increase the value of $p$. It shall be noted that $p(\bF) = 1$ when $s=2$. In addition, it is straightforward to show that $p(\bF)$ in (\ref{eq:pf1}) is greater than $1- 1/m_2^2$, and thus for a large value of $m_2$, $p(\bF)$ is very close to 1. For example, $m_2 = 11$, $p(\bF)$ is greater than 0.991.
Example \ref{ex:pf1} illustrates the use of Proposition \ref{prop:ne2} and indicates $p(\bF)$  can be very close to 1.

\begin{example}\label{ex:pf1}
Consider $s=3,\bB_{i} = \OA(81,10,3,3)$ and thus $m_2=10$,  the resulting $\bF$  is an $\OA( 243, 30, 3, 2)$ with   $p(\bF) = 0.998$ while a three-level orthogonal array
of strength three with 243 runs can accommodate up to 20 columns (Table 12.2 of \cite{HedayatSloaneStufken1999}).
Consider $s=4,\bB_{i} = \OA(64,6,4,3)$ and thus $m_2=6$,   the resulting $\bF$ is an $\OA(256,24,4,2)$ with  $p(\bF) = 0.988$ while a four-level orthogonal array of strength three with 256 runs can accommodate up to 17 columns (Table 12.6 of \cite{HedayatSloaneStufken1999}).
\end{example}

\begin{proposition}\label{prop:ne4}
If $\bA = \OA(s^2,2,s,2)$, $\bB_{i} = \OA(n_2,m_2,s,3)$, for $i=1,\ldots,s^2$ and a prime power $s$, and  $\bF$ is obtained by excluding the last two columns of $\bE$ in  (\ref{eq:E}), then  $\bF$ is an $\OA(n_2s^2, (2s-1)m_2,s,2)$ and
\begin{equation}\label{eq:pf2}
p(\bF) = 1-\frac{2m_2{s \choose 3}}{{(2s-1)m_2 \choose 3}}=1-\frac{2s(s-1)(s-2)}{(2s-1)(\gamma-1)(\gamma-2)},
\end{equation}
\noindent where $\gamma=(2s-1)m_2$.
\end{proposition}

Proposition~\ref{prop:ne4} can be shown using the same arguments as those for Proposition~\ref{prop:ne2}, with the help of Proposition \ref{prop:ne3}.  Noted that $p(\bF) = 1$ when $s=2$. For $s \geq 3$, it can be shown that
\begin{equation}
p(\bF) =
\left\{
  \begin{array}{ll}
    1-
\frac{2m_2\big [{s-1 \choose 3} +
{s-1 \choose 2}\big]}{{2(s-1)m_2+m_2 \choose 3}}, & s>3; \\
    1-
\frac{2m_2
{s-1 \choose 2}}{{2(s-1)m_2+m_2 \choose 3}}, & s=3.
  \end{array}
\right.
\end{equation}

Thus, $p(\bF)$ in (\ref{eq:pf2}) can be very close to 1 with a large value of $m_2$.
As an illustration, for $s=4,m_2=6,\bB_{i} = \OA(64,6,4,3)$,   the resulting $\bF$ of 1024 runs has 42 columns and  $p(\bF) = 0.997$ while the available four-level orthogonal array of 1024 runs can have up to 32 columns (Table 12.3 of \cite{HedayatSloaneStufken1999}).

\subsection{Resolvable orthogonal arrays}

Proposition~\ref{prop:roa} states that pairing  an orthogonal array and resolvable orthogonal arrays, the proposed method   constructs resolvable orthogonal arrays of larger sizes.

\begin{proposition}\label{prop:roa}
If $\bA=\OA(n_1,m_1,s,1)$ for $m_1=1$ or $\OA(n_1,m_1, s, 2)$ for $m_1>1$, $\bB_{i}=\ROA(n_2,m_2,s,2;\alpha)$ for $i=1,\ldots,n_1$, and a prime power  $s$, and  $\bF$ is  obtained by excluding the last $m_1$ columns of $\bE$  in (\ref{eq:E}), then 
 $\bF$
is a $\ROA(n_1n_2, (s-1)m_1m_2 + m_2, s, 2;\alpha)$.
\end{proposition}

The last $m_1$ columns of $\bE$ are excluded such that the resulting $\bF$  is the generalized Kronecker sum of
$(\bA, \bzero_{n_1})$ and the resolvable orthogonal arrays $\bB_i$'s and thus it is also resolvable. The orthogonality of $\bF$ follows by Theorem~\ref{thm:e}.
Proposition~\ref{prop:roa} offers a new way to construct resolvable orthogonal arrays which are previously obtained via the difference schemes method \cite{HedayatSloaneStufken1999}.
In addition, the result implies  an upper bound for the number of columns in a resolvable orthogonal array.
To provide the bound, we first review a result, given by \cite{Suen1989}, in Lemma~\ref{lem:cra} which says the upper
bound for  the number of columns in an $n/(sr)$-resolvable $s$-level orthogonal array is $(n-r)/(s-1)$.

\begin{lemma}\label{lem:cra}
If a resolvable $\OA(n,m,s,2)$ can be partitioned into $r$ $\OA(n/r,m,s,1)$'s, then $m \leq (n-r)/(s-1)$.
\end{lemma}

Lemma \ref{lem:cra}
says the maximal value of $m$ in a $\ROA(n,m,s,2;\alpha)$ is $[n- n/(\alpha s)]/(s-1) = n[1-1/(\alpha s)]/(s-1)$.
If in Proposition~\ref{prop:roa} we let $\bA = (0,\ldots, s-1)^T$ and   $\bB_{i} = \ROA(n_2,m_2,s,2;\alpha)$ with $m_2$
being the upper bound in  Lemma~\ref{lem:cra} for a prime power $s$, namely, $m_2 = n_2[1-1/(\alpha s)]/(s-1)$, we obtain
Proposition~\ref{prop:cra} which states that if $\bB_{i}$'s  are  $\alpha$-resolvable orthogonal arrays of $n_2$ runs and the number of
columns in each $\bB_{i}$ reaches the upper bound,  $\alpha$-resolvable orthogonal arrays of $n_2s^k$ runs with the number of columns
reaching the corresponding upper bound can be constructed, $k \geq 1$.

\begin{proposition}\label{prop:cra}
If there exists a $\ROA(n_2,m_2,s,2;\alpha)$ with $m_2 = n_2[1-1/(\alpha s)]/(s-1)$ for a prime power $s$, then
a $\ROA(n_2s^k,m,s,2;\alpha)$ with $m =  n_2s^k[1-1/(\alpha s)]/(s-1)$ and $k \geq 1$ can be constructed.
\end{proposition}

Applying Proposition~\ref{prop:cra} with $n_2=s,m_2=1,\alpha=1$,  the proposed construction in (\ref{eq:E}) provides a $\CROA(s^2,s,s,2)$ which can be used as $\bB_{i}$ to obtain a $\CROA(s^3,s^2,s,2)$, and so on.
By such repeated applications, a $\CROA(s^k,s^{k-1},s,2)$ can be constructed via  the construction   in (\ref{eq:E}).
The existing method, namely {\em difference schemes method}, can also construct a $\CROA(s^k,s^{k-1},s,2)$ \cite{HedayatSloaneStufken1999}. The advantage of the construction in (\ref{eq:E}) over the existing method is that it produces a large pool of such completely resolvable orthogonal arrays which possibly allow designs with better properties to be found.

\subsection{Balanced sliced orthogonal arrays}\label{soa}

This section shows that using a  sliced orthogonal array  for $\bA$ and an orthogonal array  for $\bB_i$, the proposed construction in  (\ref{eq:E}) provides a larger balanced sliced orthogonal array of the same levels and number of slices. Proposition \ref{prop:bsoa} provides the detailed result. Without loss of generality, the levels of orthogonal arrays in this section and next section are all labeled by the elements in $GF(s)$ and $\delta$ is a level-collapsing projection on $GF(s)$ and satisfies $\delta(a+b) = \delta(a) + \delta(b)$. The truncation projection and modulus projection defined in \cite{QianAiWu2009}, and subgroup projection defined in \cite{SunLiuQian2014} are all level-collapsing projections on $GF(s)$ and satisfy $\delta(a+b) = \delta(a) + \delta(b)$.

\begin{proposition}\label{prop:bsoa}
If $\bA = (\bA_1^T,\ldots,\bA_v^T)^T$ is an $\SOA(n_1, m_1, s, 2; v, s_0)$ where each of $\bA_1,\ldots, \bA_v$ is
an $\OA(n_0, m_1, s_0, 2)$ after some level-collapsing projection $\delta$ satisfying $\delta(a+b) = \delta(a) + \delta(b)$ for any two elements $a, b\in GF(s)$, and  for $i=1,\ldots, n_1$, $\bB_{i} = \OA(n_2,m_2,s,1)$  for $m_2=1$ and $\bB_{i} = \OA(n_2,m_2,s,2)$ for $m_2 \geq 2$, then
design $\bE$  in (\ref{eq:E})   is a $\BSOA(n_1n_2,(s-1)m_1m_2+m_1+m_2,s,2;v,s_0)$ that can be expressed as
$\bE = (\bE_1^T,\ldots,\bE_v^T)^T$ where each of $\bE_1,\ldots,\bE_v$ becomes an
$\OA(n_0n_2,(s-1)m_1m_2+m_1+m_2,s_0,2)$ after the level-collapsing projection $\delta$.
\end{proposition}

Proposition~\ref{prop:bsoa} can be shown using the same arguments as the proof of Theorem~\ref{thm:e}, and the fact that   for any design $\bD$, $\delta(\bD)$ is an $\OA(n, m, s_0, 2)$  if $\bD$ is an $\OA(n, m, s, 2)$, and under the condition $\delta(a+b) = \delta(a) + \delta(b)$, $\delta(\bD)$ is a $\D(n, m, s_0)$ if $\bD$ is a $\D(n, m, s)$.   Examples~\ref{ex:bsoa} and \ref{ex:bsoa2} illustrate the use of  the proposed construction for balanced sliced orthogonal arrays of run sizes being a prime power and not being a prime power, respectively.  

\begin{example}\label{ex:bsoa}
Consider a  $\BSOA(16,3,4,2;4,2)$ listed in Table \ref{tab:soa}. This array $\bA$ can be partitioned into $\bA_1,\ldots,\bA_4$ each of which
becomes an $\OA(4, 3, 2, 2)$ after the level collapsing $\delta$, where
$\delta(0)=\delta(x)=0, \delta(1)=\delta(x+1)=1$ and $\delta$ satisfies $\delta(a+b) = \delta(a) + \delta(b)$ for any two elements $a, b\in GF(4)$.
 Let  $\bB_i = OA(16, 5, 4,2)$ for $i=1,\ldots, 16$,
we obtain an $\bE=\BSOA(256, 53, 4, 2; 4, 2)$ that can be expressed as $\bE = (\bE_1^T,\ldots,\bE_4^T)^T$
with the property that each of $\bE_1,\ldots,\bE_4$ becomes an $\OA(64, 53, 2, 2)$ after the level-collapsing projection $\delta$. 
\end{example}

\begin{example}\label{ex:bsoa2}
Consider an $\bA=\BSOA(81,4,9,2; 9,3)$ provided by \cite{AiJiangLi2014}. This array $\bA$ can be partitioned into $\bA_1,\ldots,\bA_9$ each of which
becomes an $\OA(9, 4, 3, 2)$ after the level collapsing $\delta$, where
$\delta(0)=\delta(x+1)=\delta(2x+2)=0, \delta(1)=\delta(x+2)=\delta(2x)=1, \delta(2) = \delta(x) = \delta(2x+1)=2$ and $\delta$ satisfies $\delta(a+b) = \delta(a) + \delta(b)$ for any two elements $a, b\in GF(9)$.
 Let  $\bB_i = OA(162, 19, 9,2)$ for $i=1,\ldots, 81$,
we obtain an $\bE=\BSOA(13122, 631, 9, 2; 9, 3)$ that can be expressed as $\bE = (\bE_1^T,\ldots,\bE_9^T)^T$
with the property that each of $\bE_1,\ldots,\bE_9$ becomes an $\OA(1458, 631, 3, 2)$ after the level-collapsing projection $\delta$. 
\end{example}

\begin{table}[!ht] \begin{center}
\caption{  $\BSOA(16,3,4,2;4,2)$}
\scalebox{0.9}{
\begin{tabular}{ccccccccc}
\hline
                                 &\multicolumn{4}{c}{$\bA$}  &    &   \multicolumn{3}{c}{$\delta(\bA)$}  \\ \hline
   \multirow{4}{*}{$\bA_1$}      &   $0$     &   $0$      &     $0$   &  &  &  $0$     &   $0$      &     $0$    \\
                                 &   $1$     &   $x$      &    $x+1$  &  &  &  $1$     &   $0$      &    $1$     \\
                                 &   $x$     &   $x+1$    &     $1$   &  &  &  $0$     &   $1$      &     $1$    \\
                                 &   $x+1$   &   $1$      &    $x$    &  &  &  $1$     &   $1$      &    $0$     \\ \hline
   \multirow{4}{*}{$\bA_2$}      &   $1$     &   $1$      &     $1$   &  &  &  $1$     &   $1$      &     $1$    \\
                                 &   $0$     &   $x+1$    &    $x$    &  &  &  $0$     &   $1$      &    $0$     \\
                                 &   $x+1$   &   $x$      &     $0$   &  &  &  $1$     &   $0$      &     $0$    \\
                                 &   $x$     &   $0$      &    $x+1$  &  &  &  $0$     &   $0$      &    $1$    \\ \hline
   \multirow{4}{*}{$\bA_3$}      &   $x$     &   $x$      &     $x$   &  &  &  $0$     &   $0$      &     $0$    \\
                                 &   $x+1$   &   $0$      &    $1$    &  &  &  $1$     &   $0$      &    $1$     \\
                                 &   $0$     &   $1$      &    $x+1$  &  &  &  $0$     &   $1$      &    $1$     \\
                                 &   $1$     &   $x+1$    &    $0$    &  &  &  $1$     &   $1$      &    $0$     \\ \hline
   \multirow{4}{*}{$\bA_4$}      &   $x+1$   &    $x+1$   &   $x+1$   &  &  &  $1$     &   $1$      &   $1$      \\
                                 &   $x$     &   $1$      &    $0$    &  &  &  $0$     &   $1$      &    $0$     \\
                                 &   $1$     &   $0$      &     $x$   &  &  &  $1$     &   $0$      &     $0$    \\
                                 &   $0$     &   $x$      &    $1$    &  &  &  $0$     &   $0$      &    $1$     \\
   \hline
\end{tabular}}\label{tab:soa}
\end{center}
\end{table}

Table~\ref{tab:bsoa_list} lists the examples of balanced sliced orthogonal arrays constructed by the proposed method as in  Proposition~\ref{prop:bsoa} and by \cite{AiJiangLi2014}  without our results. Comparing $m_a$ and $m$ in Table~\ref{tab:bsoa_list},  the number of columns in  balanced sliced orthogonal arrays constructed by \cite{AiJiangLi2014} and the proposed method,  it can been seen that the proposed construction results in more columns. 

\begin{table}[!ht] \begin{center}
\caption{ Examples of balanced sliced orthogonal arrays constructed by \cite{AiJiangLi2014} and the proposed method}
\scalebox{0.9}{
\begin{tabular}{rrrrrrrrrrrr}
\hline
& $\quad n$ & $\quad  s$ & $\quad  s_0$ & $\quad v$ & $\quad  m_a$ & $\quad  m$ & $\quad n_1$ & $\quad m_1$ & $\quad n_2$ & $\quad m_2$ & 
 \\
 \hline
&64 & 4 & 2 & 4 & 12 & 13  & 16 & 3 & 4 & 1 &\\
&256 & 4 & 2 & 4 & 48 & 53  & 16 & 3 & 16 & 5 &\\
&512 & 4 & 2 & 8 & 48 & 53  & 32 & 3 & 16 & 5& \\
&1024 & 4 & 2 & 4 & 192 & 213   & 256 & 53 & 4 & 1  &\\
&512 & 8 & 2 & 8 & 56 & 57 &   64 & 7 & 8 & 1 &\\ 
&4096 & 8 & 2 & 8 & 448 & 457 &  64 &7 & 64 & 9&\\ 
&729 & 9 & 3  & 9 & 36 & 37 & 81 & 4 & 9 & 1& \\
&6561 & 9 & 3 & 3 & 243 & 253 & 81 & 3 & 81 & 10 &\\
\hline 
\multicolumn{12}{l}{\small where $m_a$ and $m$ represent the number of columns obtained by  \cite{AiJiangLi2014} and}\\
\multicolumn{12}{l}{ \small  the  proposed method, and $n_1,n_2,m_1,m_2$ are as defined in Proposition~\ref{prop:bsoa} }
\end{tabular}}\label{tab:bsoa_list}
\end{center}
\end{table}

\subsection{Nested orthogonal arrays}\label{noa}

 Proposition \ref{prop:nsoa} shows that the proposed construction in  (\ref{eq:E})  provides nested orthogonal arrays of larger run sizes using a nested orthogonal array for $\bA$ of $n_1$ runs and an orthogonal array for $\bB_i$, for $i=1,\ldots, n_1$. 
 
\begin{proposition}\label{prop:nsoa}
If $\bA = (\bA_1^T,\bA_2^T)^T$ is an $\NOA(n_1,m_1,s,2;n_0,s_0)$ where the array after applying a  level-collapsing projection $\delta$  to $\bA_1$ is an $\OA(n_0,m_1,s_0,2)$,  and $\delta$ satisfies $\delta(a+b) = \delta(a) + \delta(b)$ for any two elements $a, b\in GF(s)$, and for $i=1,\ldots,n_1$ and a prime power $s>2$, $\bB_{i} = \OA(n_2,m_2,s,1)$ for $m_2=1$ and $\bB_{i} = \OA(n_2,m_2,s,2)$ for $m_2 \geq 2$, then
design $\bE$ in  (\ref{eq:E})  is an $\NOA(n_1n_2,(s-1)m_1m_2+m_1+m_2,s,2;n_0n_2,s_0)$ that can be expressed as $\bE = (\bE_1^T,\bE_2^T)^T$ where $\bE_1$ becomes an $\OA(n_0n_2,(s-1)m_1m_2+m_1+m_2,s_0,2)$ after the level-collapsing  projection $\delta$.
\end{proposition}

The proof of Proposition \ref{prop:nsoa} is similar to that of Theorem \ref{thm:e} and thus we omit it.  Example \ref{ex:noa} gives an illustration of the use of Proposition \ref{prop:nsoa}.

\begin{example}\label{ex:noa}
Let $\bA$ be the $\OA(16,3,4,2)$ listed in  Table \ref{tab:soa} and $\delta$ defined in Example \ref{ex:bsoa}. It is easy to check $\delta(\bA_1)$  being an $\OA(4, 3, 2, 2)$, thus  $\bA=(\bA^T_1, \ (\bA^T_2, \bA^T_3, \bA^T_4))^T$ is an $\NOA(16, 3, 4, 2; 4, 2)$. For $i=1, \ldots, 4$, taking $\bB_i = \OA(16, 5, 4,2)$ from the orthogonal array website \cite{Sloane2004}, we obtain an $\NOA(256, 53, 4, 2; 64, 2)$ based on Proposition \ref{prop:nsoa}, and the array has more columns that an $\NOA(256, 48, 4, 2; 64, 2)$ provided by \cite{QianAiWu2009}. Another example is,  using an $\NOA(64,5,8,2;16,4)$ and an $\OA(64,9,8,2)$, we have an $\NOA(4096, 329,8,2;1024,4)$. However, \cite{QianAiWu2009} can only give an $\NOA(4096, 320,8,2;1024,4)$ with an $\NOA(64,5,8,2;16,4)$ and a $\D(64,64,8)$. 
\end{example}

\section{Conclusion and Discussion}

We introduce a versatile method for constructing different kinds of orthogonal arrays. Armed with the theoretical results in Sections 3 and 4, we demonstrate that the proposed construction provides countless series of orthogonal arrays of strength two and three. Some examples of such arrays are listed in Table~\ref{table:oa} and Example~\ref{ex:oa3_243}. Examples of new sliced orthogonal arrays and nested orthogonal arrays are given in Table~\ref{tab:bsoa_list} and Example~\ref{ex:noa}.  It shall be emphasized that, although the proposed method constructs nearly the same orthogonal arrays of strength two and three as the existing approaches in terms  of the run sizes and the number of factors, it provides new orthogonal arrays of strength two and three by applying the proposed structure using non-isomorphic $\bA$'s, non-isomorphic $\bB_i$'s or isomorphic $\bB_i$'s up to column permutations and/or level relabeling of one or more columns.  The same statement applies to resolvable orthogonal arrays.  Particularly, for orthogonal arrays of large run sizes, the proposed method provides a new, flexible and efficient way to find good designs. For complete results on new orthogonal arrays of strength two and three, resolvable orthogonal arrays, sliced orthogonal arrays and nested orthogonal arrays offered by the proposed construction, it deserves a comprehensive investigation and thus we do not dwell on this issue here. 

The proposed method is shown to construct balanced sliced orthogonal arrays,  nested orthogonal arrays, and 
orthogonal arrays of (near) strength three, the last of which is a new class of orthogonal arrays we introduce.  Compared with the existing approaches for constructing balanced sliced orthogonal arrays and nested orthogonal arrays, the proposed method provides new such arrays of run sizes that are not prime powers, noting that  \cite{AiJiangLi2014} only constructs the balanced sliced orthogonal arrays of runs sizes being prime powers, and \cite{QianTangWu2009} and \cite{SunLiuQian2014} only offer the nested orthogonal arrays of run sizes being prime powers.

One direction for future work is to apply the proposed approach to obtain optimal orthogonal arrays in a similar fashion as in \cite{XuChengWu2004}
which combines the construction with the algorithmic procedure. Another direction is to use this structure to obtain the catalogue of optimal designs
\cite{LinSitterTang2012}.  One direction that deserves exploration is the use of the proposed approach for constructing mixed-level orthogonal arrays, and its use in constructing strong orthogonal arrays \cite{HeTang2013, ZhouTang2019}.

%%%%%%%%%%%%%%%%%%%%%%%%%%%%%%%%%%%%%%%%%%%%%%
%% Example with single Appendix:            %%
%%%%%%%%%%%%%%%%%%%%%%%%%%%%%%%%%%%%%%%%%%%%%%
\begin{appendix}
\section*{Proofs}\label{appn} %% if no title is needed, leave empty \section*{}.

\noindent {\em Proof of Lemma~\ref{lem:2}}:
 It is easy to obtain items $(i)$ and $(iii)$.
                 Under item $(ii)$, by row permutations, $({\bf a}\oast{\bf b}, {\bf c}\oast\alpha{\bf b})$ can become $n_2/s$ repetitions of
                 $(\xi\oplus{\bf a}, \alpha\xi\oplus{\bf c})$, where $\xi=(\alpha_0, \alpha_1, \ldots, \alpha_{s-1})^T$. Then
                 items $(ii.1)$ to $(ii.3)$ follow from this structure. Under item $(ii.4)$, note that $(\xi, \alpha\xi)$ is a $D(s, 2, s)$ when $\alpha\neq 1$. In addition, as ${\bf a}$ is balanced, then $(\xi\oplus{\bf a}, \alpha\xi\oplus{\bf a})$ is an orthogonal array.
                 Hence the conclusion is true.

\vspace{4mm}
\noindent {\em Proof of Lemma~\ref{lem:3}:}
Item $(i)$ is obvious according to the definition of generalized Kronecker sum and orthogonal arrays of strength three. Item $(iv)$ is directly followed by Lemma~\ref{lem:2}. We show items $(ii)$ and $(iii)$ respectively. For items $(ii)$ and $(iii.1)$ and $(iii.2)$, we prove that
for any level combination $(\alpha^*, \beta^*)$ in the rows of $({\bf a}\oast {\bf b}, {\bf c}\oast {\bf d})$, the corresponding
rows in ${\bf e}\oast {\bf f}$ form a balanced design. Under item $(ii)$, since $({\bf b}_i, {\bf d}_i)=OA(n_2, 2, s, 2)$,  for any pair of
$(\alpha^*, \beta^*)$, there are $n_2/s^2$ rows in $(a_i+{\bf b}_i, c_i+{\bf d}_i)$ equal to $(\alpha^*, \beta^*)$, and for the corresponding
rows in $e_i+{\bf f}_i=e_i+\alpha{\bf b}_i$ take the value of $\alpha\alpha^*+(e_i-\alpha a_i)$. Because
$({\bf e}, \alpha{\bf a})=D(n_1, 2, s)$, that implies $\{\alpha\alpha^*+(e_i-\alpha a_i)\}_{i=1}^{n_1}$
take each value of $GF(s)$ with the same frequency, and hence those rows in ${\bf e}\oast{\bf f}$ are balanced. Under item $(iii)$, as $({\bf a}, {\bf c})=OA(n_1, 2, s, 2)$, we have $({\bf a}\oast {\bf b}, {\bf c}\oast \beta{\bf b})$ is
an $OA(n_1n_2, 2, s, 2)$ according to item $(ii.2)$ of Lemma~\ref{lem:2}, and for the rows in
$(a_i + {\bf b}_i, c_i +\beta{{\bf b}_i})$ equal to $(\alpha^*, \beta^*)$, we have $c_i-\beta a_i=\beta^*-\beta\alpha^*$.
If $\beta=0$, it is clear that there are $n_1/s$ rows in $(\beta \ba, \bc)$ satisfying the relationship $c_i - \beta a_i = \beta^*- \beta a^*$. If $\beta \neq 0$,  as $({\beta{\bf a}, {\bf c}})=OA(n_1, 2, s, 2)$,  there are $n_1/s$ rows in
$({\beta \bf a}, {\bf c})$  satisfying the relationship  $c_i-\beta a_i=\beta^*-\beta\alpha^*$, and both ${\bf a}$ and ${\bf c}$ are still balanced in such $n_1/s$ rows.
Correspondingly, in ${\bf e}\oast{\bf f}$, we have $e_i + \alpha {\bf b}_i=e_i + \alpha(\alpha^*-a_i)=\alpha\alpha^*+(e_i-\alpha a_i)$  as $b_i = \alpha^*-a_i$.
Under item $(iii.1)$, $e_i-\alpha a_i=(1-\alpha)a_i$; or under item $(iii.2)$, $e_i-\alpha a_i=-\alpha a_i$, both are balanced in the corresponding $n_1/s$ rows.
Therefore, either in item $(iii.1)$ or $(iii.2)$, the corresponding rows in ${\bf e}\oast {\bf f}$ are balanced. Thus, items $(iii.1)$ and $(iii.2)$ follow. Under item $(iii.3)$,
consider $({\bf a}\oast {\bf b}, {\bf c}\oast \beta{\bf b}, {\bf e}\oast\alpha{\bf b})$.
By row permutations, the array can be arranged as $n_2/s$ repetitions of $(\xi\oplus{\bf a}, \beta\xi\oplus{\bf c}, \alpha\xi\oplus{\bf e})$,
which is an $OA(sn_1, 3, s, 3)$ by $({\bf a}, {\bf c}, {\bf e})=OA(n_1, 3, s, 3)$, where $\xi=(\alpha_0, \alpha_1, \ldots, \alpha_{s-1})^T$.
Thus, item $(iii.3)$ follows. We complete the proof.

\vspace{4mm}
\noindent {\em Proof of Theorem~\ref{thm:e}:}
Theorem~\ref{thm:e} can be verified by checking the orthogonality of any two columns in $\bE$ in  (\ref{eq:E}) according to Lemma~\ref{lem:2}. More specifically, consider any two columns of $\bE$, which can be represented by $\bu = \ba_p \oast (\alpha_j * \bb_q)$ and $\bv = \ba_{p'} \oast (\alpha_{j'}*\bb_{q'})$, Table~\ref{table:T1} lists conditions in Lemma~\ref{lem:2} used for all combinations of $\bu$ and $\bv$. Note that $g=1,\ldots,s-1$ in the table.
\begin{table}[h]
\begin{center}
\caption{Conditions in Lemma~\ref{lem:2} used for different $\bu$ and $\bv$}
\begin{tabular}{l l l l  l}
\hline
$\bu$ and $\bv$ &$\quad$ &$(p,q)$ and $(p',q')$ & $\quad$ & Condition \\
\hline
\multirow{2}{*}{$\bu \in D_g$ and $\bv \in D_g$}
&& $q \neq q'$ && $(i)$\\
&& $p\neq p',q = q'$ && $(ii.2)$\\
\hline
$\bu \in D_s, \bv \in D_s$ & & $q \neq q'$ && $(i)$\\
\hline
$\bu \in \bA \oplus \bzero_{n_2}, \bv \in \bA \oplus \bzero_{n_2}$
&& $p \neq p'$ && $(iii.2)$ \\
\hline
\multirow{3}{*}{$\bu \in D_g$ and $\bv \in D_{g'}$, $g \neq g'$}
&& $q \neq q'$ && $(i)$\\
&& $p\neq p',q = q'$ && $(ii.2)$\\
&& $p=p',q = q'$ && $(ii.4)$\\
\hline
\multirow{2}{*}{$\bu \in D_g$ and $\bv \in D_s$}
&& $q \neq q'$ && $(i)$\\
&& $q = q'$ && $(ii.3)$\\
\hline
$\bu \in D_g, \bv \in \bA \oplus \bzero_{n_2}$
&&  &&$(iii.1)$\\
\hline
$\bu \in D_s, \bv \in \bA \oplus \bzero_{n_2}$
&& && $(iii.1)$\\
\hline
\end{tabular}
\end{center}
\label{table:T1}
\end{table}

\end{appendix}
%%%%%%%%%%%%%%%%%%%%%%%%%%%%%%%%%%%%%%%%%%%%%%
%% Example with multiple Appendixes:        %%
%%%%%%%%%%%%%%%%%%%%%%%%%%%%%%%%%%%%%%%%%%%%%%
%\begin{appendix}
%\section{Title of the first appendix}\label{appA}
%If there are more than one appendix, then please refer to it
%as \ldots\ in Appendix \ref{appA}, Appendix \ref{appB}, etc.

%\section{Title of the second appendix}\label{appB}
%\subsection{First subsection of Appendix \protect\ref{appB}}

%Use the standard \LaTeX\ commands for headings in \verb|{appendix}|.
%Headings and other objects will be numbered automatically.
%\begin{equation}
%\mathcal{P}=(j_{k,1},j_{k,2},\dots,j_{k,m(k)}). \label{path}
%\end{equation}

%Sample of cross-reference to the formula (\ref{path}) in Appendix \ref{appB}.
%\end{appendix}

%%%%%%%%%%%%%%%%%%%%%%%%%%%%%%%%%%%%%%%%%%%%%%
%% Support information, if any,             %%
%% should be provided in the                %%
%% Acknowledgements section.                %%
%%%%%%%%%%%%%%%%%%%%%%%%%%%%%%%%%%%%%%%%%%%%%%
\begin{acks}[Acknowledgments]
We extend our sincere thanks to the Editor, Associate Editor and several
referees for their constructive comments that lead to numerous improvements over the previous versions.
\end{acks}

%%%%%%%%%%%%%%%%%%%%%%%%%%%%%%%%%%%%%%%%%%%%%%
%% Funding information, if any,             %%
%% should be provided in the                %%
%% funding section.                         %%
%%%%%%%%%%%%%%%%%%%%%%%%%%%%%%%%%%%%%%%%%%%%%%
\begin{funding}
He was supported by   National Natural Science Foundation of China, Grant No.\ 11701033.
Lin was supported by Discovery grant from the Natural Sciences and Engineering Research Council of Canada. Sun is supported by Supported by National Natural Science Foundation of China Grant No.\ 11971098 and National Key Research and Development Program of China (No. 2020YFA0714102).\end{funding}

%%%%%%%%%%%%%%%%%%%%%%%%%%%%%%%%%%%%%%%%%%%%%%
%% Supplementary Material, including data   %%
%% sets and code, should be provided in     %%
%% {supplement} environment with title      %%
%% and short description. It cannot be      %%
%% available exclusively as external link.  %%
%% All Supplementary Material must be       %%
%% available to the reader on Project       %%
%% Euclid with the published article.       %%
%%%%%%%%%%%%%%%%%%%%%%%%%%%%%%%%%%%%%%%%%%%%%%
\begin{supplement}
\stitle{Supplement to ``A New and Flexible Design Construction for Orthogonal Arrays for Modern Applications''}
\sdescription{(doi: COMPLETED BY THE TYPESETTER.pdf). We provide results on determining  the maximal number of columns in $\bE$ in  (\ref{eq:E})  for given values of $n$, $s$ and $t$, and new results on  lower bounds of the maximal number of columns in certain orthogonal arrays of strength three. 
}
\end{supplement}

%\begin{supplement}
%\stitle{Title of Supplement B}
%\sdescription{Short description of Supplement B.}
%\end{supplement}

%%%%%%%%%%%%%%%%%%%%%%%%%%%%%%%%%%%%%%%%%%%%%%%%%%%%%%%%%%%%%
%%                  The Bibliography                       %%
%%                                                         %%
%%  imsart-???.bst  will be used to                        %%
%%  create a .BBL file for submission.                     %%
%%                                                         %%
%%  Note that the displayed Bibliography will not          %%
%%  necessarily be rendered by Latex exactly as specified  %%
%%  in the online Instructions for Authors.                %%
%%                                                         %%
%%  MR numbers will be added by VTeX.                      %%
%%                                                         %%
%%  Use \cite{...} to cite references in text.             %%
%%                                                         %%
%%%%%%%%%%%%%%%%%%%%%%%%%%%%%%%%%%%%%%%%%%%%%%%%%%%%%%%%%%%%%

%% if your bibliography is in bibtex format, uncomment commands:
%\bibliographystyle{imsart-number} % Style BST file (imsart-number.bst or imsart-nameyear.bst)
%\bibliography{bibliography}       % Bibliography file (usually '*.bib')

%% or include bibliography directly:

\end{document}